\documentclass[reprint,showpacs,amsmath,amssymb,pra]{revtex4-1}
\usepackage{graphicx}
\usepackage{dcolumn}
\usepackage{bm}
\usepackage{color}

\begin{document}

\title{Transmission enhancement in loss-gain multilayers by resonant suppression of reflection}
%\title{Light propagation in active metal-dielectric multilayers: Absorption compensation and resonant low-reflection transmission}

\author{Denis~V.~Novitsky$^{1,2}$}
\email{dvnovitsky@gmail.com}
\author{Vladimir~R.~Tuz$^{3,4}$}
\author{Sergey~L.~Prosvirnin$^{4,5}$}
\author{Andrei~V.~Lavrinenko$^{6}$}
\author{Andrey~V.~Novitsky$^{6,7}$}
\email{anov@fotonik.dtu.dk} \affiliation{$^1$B. I.
Stepanov Institute of Physics, National Academy of Sciences of
Belarus, 68 Nezavisimosti Avenue,
Minsk 220072, Belarus\\
$^2$ITMO University, 49 Kronverksky Pr., St. Petersburg 197101, Russia\\
$^3$International Center of Future Science, State Key Laboratory on Integrated Optoelectronics, College of Electronic Science and Engineering, Jilin University, 2699  Qianjin Str., Changchun 130012, China\\
$^4$Institute of Radio Astronomy of National Academy of Sciences of
Ukraine, 4 Mystetstv Street, Kharkiv 61002, Ukraine\\
$^5$School of Radio Physics, V.N. Karazin Kharkiv National
University, 4 Svobody sq., Kharkiv 61022, Ukraine\\
$^6$DTU Fotonik, Technical University of Denmark, {\O}rsteds Plads
343, DK-2800 Kongens Lyngby, Denmark\\
$^7$Department of Theoretical Physics and Astrophysics, Belarusian
State University, 4 Nezavisimosti Avenue, Minsk 220030, Belarus}

\date{\today}

\begin{abstract}
% Plasmonic devices, in spite of their fascinating properties for controlling light at nanoscale, possess inevitable losses.
Using the transfer-matrix approach and solving time-domain
differential equations, we analyze the loss compensation mechanism
in multilayer systems composed of an absorbing transparent
conductive oxide and dielectric doped with an active material. We
reveal also another regime with the possibility of enhanced
transmission with suppressed reflection originating from the
resonant properties of the multilayers. For obliquely incident and
evanescent waves, such enhanced transmission under suppressed
reflection turns into the reflectionless regime, which is similar to
that observed in the $\mathcal{PT}$-symmetric structures, but does
not require $\mathcal{PT}$ symmetry. We infer that the
reflectionless transmission is due to the full loss compensation at
the resonant wavelengths of the multilayers.
%We envisage applications of resonant enhanced transmission for suppressing losses in plasmonic and metamaterial structures.
\end{abstract}

\maketitle

\section{Introduction}

Metals play a tremendous role in modern optics and nanophotonics due
to their ability for light concentration in tiny volumes
\cite{Novotny}. Field enhancement is realized through excitation of
propagating and/or localized surface plasmons -- and can be used for
strengthening linear and nonlinear light-matter interactions
\cite{Meier,Bozhevolnyi}. However, electromagnetic waves in
structures containing metals suffer from losses, and the beneficial
plasmonic properties may fade away. One of the simplest transmission
systems is a one-dimensional (1D) metal-dielectric multilayer
structure. It attracts much attention nowadays, since it provides
one of the possible implementations of so-called hyperbolic
metamaterials (HMMs) \cite{Poddubny,Drachev}. HMMs are able to
support waves with high wavenumbers, which are typically evanescent
in homogeneous dielectrics, thus allowing superresolution, the
Purcell factor enhancement, and so on \cite{Lu,Ferrari,Cortes}.
However, attenuation of transmitted waves greatly degrades the
performance of the HMM-based devices.

Transparent conductive oxides (TCOs), such as tin-doped indium-oxide
(ITO), aluminum-doped zinc-oxide (AZO), and gallium-doped zinc-oxide
(GZO), exhibit reasonably low losses and, thus, can be employed as
alternative plasmonic materials
\cite{Naik11,Naik13,Babicheva15,Guo16,Wang17}. Another simple recipe
for mitigating the absorption problem is loss compensation using
amplifying materials. Optical gain can be delivered by
low-dimensional electronic systems such as quantum dots and quantum
wells, active impurities, or new phases in the semiconductor
lattice, as well as by bulk semiconductors themselves \cite{Pavesi}.
In the case of atomically thin transition metal dichalcogenides the
optical gain can be enhanced using the nanostructured plasmonic
substrate as demonstrated in Ref. \cite{Jeong16}. Using active
slabs, losses can be compensated in surface plasmonic waveguides
\cite{Zayats13,Oulton09,Svintsov} and negative-refractive-index
metamaterials \cite{Zhang08,Xiao10} under optical and electrical
pumping. Loss compensation in metal-dielectric multilayers is mainly
directed towards the hyperbolic metamaterials improvement
\cite{Ni11,Savelev,Argyropoulos} and can be used for enhancement of
spontaneous emission \cite{Galfsky} ameliorating of surface plasmon
resonances  \cite{Pustovit}, as well as for tuning gain/absorption
spectra in graphene HMMs \cite{Janaszek}.

The losses in coalescence with gain lead to the remarkable property
of the multilayers known as the parity-time ($\mathcal{PT}$)
symmetry. Optical $\mathcal{PT}$ symmetry is transferred from
quantum mechanics of non-Hermitian Hamiltonians \cite{Bender98}. It
can be realized by the sequencing of active and passive structural
components, which for the complex 1D permittivity reads
$\varepsilon(z) = \varepsilon^\ast(-z)$ (here $^\ast$ stands for the
complex conjugate). $\mathcal{PT}$ symmetry opens new opportunities
for light control \cite{ElGanainy07,Ruter10} due to nonreciprocity
of light propagation \cite{Makris08}, anisotropic transmission
resonances \cite{Ge12}, and unidirectional invisibility
\cite{Lin11}. It influences light localization in disordered
structures \cite{Kartashov16} and manifests new regularities of
optical switching and soliton generation in nonlinear systems
\cite{Suchkov16,Konotop16}.

In most of the theoretical works gain in multilayers is modeled
phenomenologically through the imaginary part of the complex
permittivity. However, realistic modeling should include the
temporal dynamics of light interaction with the amplifying medium.
Only within the dynamical framework one can describe the fast
processes and influence of the inhomogeneous field distribution
inside a gain slab, as well as correctly predict establishment of
the stationary state or lasing in metal-dielectric structures
\cite{Fietz12,Ruting14,Veltri16}. For the multilayer we reveal the
frames of applicability of the transfer-matrix method using
comparison with the in-house finite-difference time-domain (FDTD)
method \cite{Tuz2014,Novit2009}. The FDTD approach describes the
frequency-dispersive metal-like material response with the
time-dependent equations, while the Bloch equations are employed for
dielectric having resonant gain. Here we combine the advantages of
the loss suppression in TCOs (AZO) and amplification in a gain
material (with, i.e., quantum dots).

The paper is organized as follows. In Sec. \ref{eqpars}, we discuss
the main equations governing light propagation in metal-dielectric
multilayers. Section \ref{compHMM} delivers a full-temporal evidence
of loss compensation and signal amplification in active systems with
and without saturation. The effect of resonant low-reflection
transmission is introduced in Sec. \ref{trans} on the basis of
calculations with both the FDTD and transfer-matrix techniques.
Development of the lasing-like behavior, limiting the applicability
of the transfer-matrix method, is studied as well. The resonant
reflectionless transmission is observed in the case of obliquely
incident light and evanescent waves, but the electromagnetic states
differ from those of the $\mathcal{PT}$-symmetric systems as shown
in Sec. \ref{oblique}. Section \ref{concl} summarizes the article.

\section{\label{eqpars} Modeling loss and gain in time domain}

\begin{figure}[t!]
\includegraphics[scale=0.6, clip=]{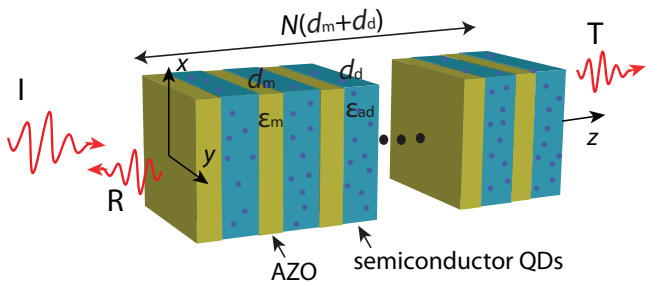}
\caption{\label{fig1} Sketch of the system under investigation: $N$
periods of the alternating lossy AZO and amplifying dielectric
slabs.}
\end{figure}

We consider a periodic planar structure composed
of alternating metal-like and dielectric layers (Fig.~\ref{fig1}) and
illuminated by normally incident monochromatic light [convention $\exp(-i \omega t)$ is assumed, where $\omega$ is the angular frequency]. The permittivity of the metal-like slabs can be written according to the Drude model as
\begin{eqnarray}
\varepsilon_m(\omega)=\varepsilon_\infty - \frac{\omega^2_p}{\omega^2+i \gamma \omega},
\label{epsm}
\end{eqnarray}
where $\varepsilon_\infty$ is the high-frequency dielectric
constant, $\omega_p$ is the plasma frequency, and $\gamma$ is the
electron plasma damping. Since we describe light propagation in
the metal-dielectric structure in the time domain, Eq.~(\ref{epsm}) should be converted into a differential equation with respect to time. Introducing derivative $\partial/\partial t$ instead of $-i \omega$, we straightforwardly arrive at the oscillation equation for polarization
\begin{eqnarray}
\frac{d^2 P}{dt^2}+\gamma \frac{d P}{dt}=\omega^2_p E. \label{polmet}
\end{eqnarray}
From here on we assume that the wave is linearly polarized, and we
can use the scalar quantities of electric field $E$ and polarization
$P$. Then light dynamics in the metal-like material is governed by the wave
equation
\begin{eqnarray}
\frac{\partial^2 E}{\partial z^2} -
\frac{\varepsilon_\infty}{c^2}\frac{\partial^2 E}{\partial t^2} =
\frac{1}{c^2}\frac{\partial^2 P}{\partial t^2}. \label{weqmet}
\end{eqnarray}
Having carrier frequency $\omega$, we represent polarization
$P=p(t,z) \exp{[-i(\omega t-kz)]}$ and electric field $E=A(t,z)
\exp{[-i(\omega t-kz)]}$ in terms of the complex amplitudes $p(t,z)$
and $A(t,z)$. Then Eqs.~(\ref{polmet}) and (\ref{weqmet}) read as
\begin{eqnarray}
\frac{d^2 p}{d\tau^2}&+&\eta \frac{dp}{d\tau}+\zeta p=\alpha A,
\label{polmet1} \\
\frac{\partial^2 A}{\partial \xi^2}&-& \varepsilon_\infty
\frac{\partial^2 A}{\partial \tau^2}+2 i \frac{\partial A}{\partial
\xi}+2 i \varepsilon_\infty \frac{\partial
A}{\partial \tau} + A (\varepsilon_\infty-1) \nonumber \\
&&= \frac{\partial^2 p}{\partial \tau^2}-2 i \frac{\partial
p}{\partial \tau}-p, \label{weqmet1}
\end{eqnarray}
where $\tau=\omega t$ and $\xi=kz$ are the dimensionless time and
distance, $k=\omega /c$ is the vacuum wavenumber, $c$ is the speed
of light in free space, $\eta=-2i+\gamma/\omega$,
$\zeta=-1-i\gamma/\omega$, and $\alpha=\omega^2_p/\omega^2$.
Equations~(\ref{polmet1}) and (\ref{weqmet1}) can be numerically
solved using the FDTD approach proposed earlier for the examination
of nonlinear metamaterials \cite{Tuz2014}.

The second component of the structure is an active dielectric (gain
medium) modeled as a homogeneously broadened two-level medium. Light
propagates in this medium in concordance with the Maxwell-Bloch
equations, that is, the system of differential equations for the
dimensionless electric-field amplitude $\Omega=(\mu/\hbar \omega) A$
(normalized Rabi frequency), complex amplitude of the atomic
polarization $\rho$, and difference between populations of ground
and excited states $w$ \cite{Novit2011}:
\begin{eqnarray}
\frac{d\rho}{d\tau}&=& i l \Omega w + i \rho \delta - \gamma_2 \rho, \label{dPdtau} \\
\frac{dw}{d\tau}&=&2 i (l^* \Omega^* \rho - \rho^* l \Omega) -
\gamma_1 (w-w_{eq}),
\label{dNdtau} \\
\frac{\partial^2 \Omega}{\partial \xi^2}&-& n_d^2 \frac{\partial^2
\Omega}{\partial \tau^2}+2 i \frac{\partial \Omega}{\partial \xi}+2
i n_d^2 \frac{\partial \Omega}{\partial
\tau} + (n_d^2-1) \Omega \nonumber \\
&&=3 \epsilon l \left(\frac{\partial^2 \rho}{\partial \tau^2}-2 i
\frac{\partial \rho}{\partial \tau}-\rho\right), \label{Maxdl}
\end{eqnarray}
where $\mu$ is the dipole moment of the quantum transition, $\hbar$
is the reduced Planck constant, $\delta=\Delta
\omega/\omega=(\omega_0-\omega)/\omega$ is the normalized frequency
detuning, $\omega_0$ is the frequency of the atomic resonance,
$\gamma_{1}=1/(\omega T_{1})$ and $\gamma_{2}=1/(\omega T_{2})$ are
respectively normalized relaxation rates of population and
polarization, $T_1$ ($T_2$) is the longitudinal (transverse)
relaxation time, and $w_{eq}$ is the equilibrium population
difference. Light-matter coupling is described by the dimensionless
parameter $\epsilon= \omega_L / \omega = 4 \pi \mu^2 C/3 \hbar
\omega$, where $\omega_L$ is the normalized Lorentz frequency and
$C$ is the concentration (density) of two-level atoms. Quantity
$l=(n_d^2+2)/3$ is the so-called local-field enhancement factor
originating from polarization of the background dielectric with
refractive index $n_d$ by the embedded active (two-level) particles
\cite{Crenshaw, Novit2011a}. Further we numerically solve Eqs.
(\ref{dPdtau})--(\ref{Maxdl}) using essentially the same FDTD
approach as in our previous publications (e.g., see
\cite{Novit2009}). It is worth noting that at the edges of the
calculation region we apply perfectly matched layers (PMLs) and use
the total field / scattered field (TF/SF) method to set the incident
field. Solutions of both linear equations (\ref{polmet1}) and
(\ref{weqmet1}) in conductive layers and nonlinear equations
(\ref{dPdtau})--(\ref{Maxdl}) in dielectric layers yield the
temporal evolution of light in the whole structure.

Equilibrium population difference $w_{eq}$ is the key parameter
governing the gain properties of dielectric. Without loss of
generality it can be considered as a parameter of the pumping
intensity. Indeed, when there is no external pump, all two-level
atoms are in the ground state, i.e., the difference between the
populations of the ground and excited states is $w_{eq}=1$. In the
opposite case of totally inverted medium, all atoms are excited by
the pump and $w_{eq}=-1$. The medium is saturated, if both levels
are populated equally in the state of equilibrium. This gainless
case corresponds to $w_{eq}=0$ being equivalent to the absence of
transitions between the levels; in other words, the absence of
two-level atoms themselves. We are apparently interested in nonzero
inversion, which can serve as a simple model of the gain medium.

At first, we will demonstrate the link between the equilibrium
population difference and gain estimating the permittivity of the
two-level medium in the steady-state approximation. Eliminating the
time derivatives in Eqs. (\ref{dPdtau}) and (\ref{dNdtau}), we get
the stationary microscopic polarization $\rho_{st}$ and population
difference $w_{st}$ as follows:
\begin{eqnarray}
\rho_{st} &=& \frac{i \gamma_2-\delta}{\gamma_2^2+\delta^2}\frac{lw_{eq}}{1+|\Omega|^2/\Omega^2_{sat}} \Omega, \label{pstat} \\
w_{st} &=& \frac{w_{eq}}{1+|\Omega|^2/\Omega^2_{sat}}, \label{Nstat}
\end{eqnarray}
where $\Omega_{sat}=\sqrt{\gamma_1 (\gamma_2^2+\delta^2)/4l^2
\gamma_2}$ is the Rabi frequency corresponding to the saturation
intensity. Such a medium with saturable nonlinearity is
characterized by the effective permittivity
\begin{eqnarray}
\varepsilon_{ad} &=& n_d^2+4 \pi \mu C \rho_{st}/E
= n_d^2+\frac{K(-\delta+i\gamma_2)}{1+|\Omega|^2/\Omega^2_{sat}},
\label{epsTLM}
\end{eqnarray}
where $K=3 \omega_L l^2 w_{eq}/[\omega (\gamma_2^2+\delta^2)]$.
According to Ref. \cite{Crenshaw}, additional enhancement of the
resonant polarization is taken into account. Assuming that $n_d$ is
real, the imaginary part of complex permittivity $\varepsilon_{ad}$
is a function of pumping parameter $w_{eq}$. In order to achieve a
net gain, the pumping parameter should be obviously negative.
Applying further simplifications of the exact resonance $\delta=0$
and low-intensity external radiation $|\Omega| \ll\Omega_{sat}$, Eq.
(\ref{epsTLM}) transforms to
\begin{eqnarray}
\varepsilon_{ad} \approx n_d^2+3 i l^2 \omega_L T_2 w_{eq}.
\label{epsTLMsimple}
\end{eqnarray}
This formula will be exploited further for calculations of the
transmission and reflection spectra of the multilayers within the
transfer-matrix method \cite{Markos}. It is worth noting that
according to Eq. (\ref{epsTLMsimple}) the imaginary addition to
permittivity is influenced only by the transverse (coherence)
relaxation time $T_2$. The longitudinal (population) relaxation time
$T_1$ defines the saturation intensity that limits the validity
range of Eq.~(\ref{epsTLMsimple}).

In this paper, we use Al-doped ZnO (AZO) as a conducting material.
According to optical characterization of AZO films \cite{AZO}, this
material can be described by Eq. (\ref{epsm}) with
$\varepsilon_\infty = 3.45$, $\omega_p = 2 \pi \times 274.5$ THz,
and $\gamma = 2 \pi \times 35$ THz. As an active dielectric, we
choose a semiconductor doped with quantum dots at the exact
resonance having the following parameters \cite{Palik,Diels}:
$n_d=3.3$, $\omega_L=10^{11}$ s$^{-1}$, $T_1=1$ ns, and $T_2=0.1$
ps. Estimation of the permittivity imaginary part according to Eq.
(\ref{epsTLMsimple}) yields $|\textrm{Im} \varepsilon_{ad}| \lesssim
0.5$ at $w_{eq}=-1$. The corresponding gain coefficient $g=4 \pi
\textrm{Im}(\sqrt{\varepsilon_{ad}})/\lambda \lesssim 5 \cdot 10^3$
cm$^{-1}$ at wavelength $\lambda \sim 2$ $\mu$m is indeed attainable
with current semiconductor technologies \cite{Savoia,Babicheva12}.
It is also worth noting that the choice of materials is not unique
and the similar results could be obtained for other sets of
materials, if geometry of the structure and light wavelengths are
appropriately adjusted.

Hereinafter, normalized Rabi frequency $\Omega = \mu A/\hbar \omega$
is used instead of the electric field amplitude $A$. The values of
$\Omega$ are written in the units of dimensionless relaxation rate
$\gamma_2$. Throughout the text of the paper the initial value of
the population difference is equal to the pumping parameter as
$w(t=0)=w_{eq}$.

\section{\label{compHMM} Absorption compensation in lossy multilayers}

\begin{figure}[t!]
\includegraphics[scale=0.85, clip=]{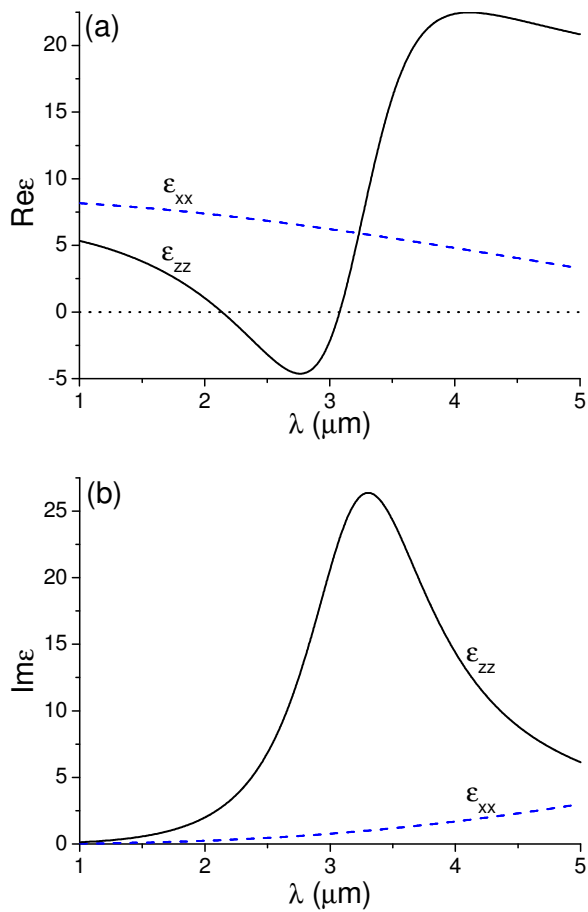}
\caption{\label{fig2} (a) Real and (b) imaginary parts of the
effective permittivity tensor components of the
AZO-dielectric multilayer ($N = 50$) structure. Parameters of the
AZO slab: $\varepsilon_\infty = 3.45$, $\omega_p = 2 \pi \times
147.8$ THz, $\gamma = 2 \pi \times 35$ THz, and $d_m=0.1$ $\mu$m.
Parameters of the gainless ($w_{eq} = 0$) dielectric slab:
$n_d=3.3$, $\omega_L=10^{11}$ s$^{-1}$, $T_1=1$ ns, and $T_2=0.1$
ps, and $d_m=0.2$ $\mu$m.}
\end{figure}

\begin{figure*}[t!]
\includegraphics[scale=0.85, clip=]{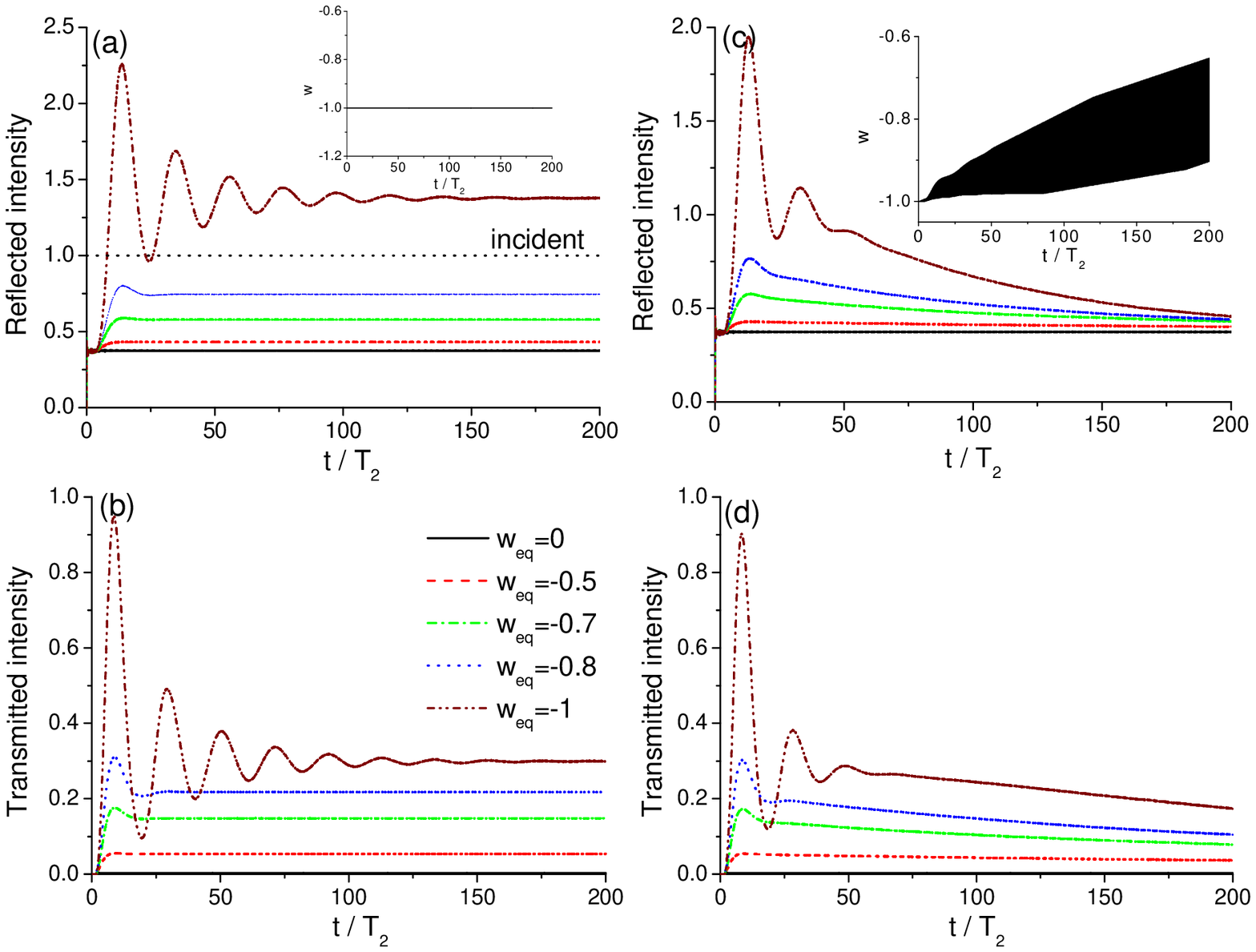}
\caption{\label{fig3} Temporal dynamics of (a),(c) reflected and
(b),(d) transmitted intensities for different $w_{eq}$ provided the
cw input has the amplitude (a),(b) $\Omega_0 = 10^{-4} \gamma_2$ and
(c),(d) $\Omega_0 = 10^{-2} \gamma_2$. Dynamics of the population
difference at the entrance of the first gain layer for each of two
input amplitudes $\Omega_0$ and pumping parameter $w_{eq}=-1$ are
shown in the insets of (a) and (c). Wavelength equals $\lambda=2.5$
$\mu$m; other parameters are the same as in Fig. \ref{fig2}.}
\end{figure*}

Let us apply the model described above to investigation of loss
compensation in periodic planar metamaterials composed of
alternating layers of AZO and dielectric with gain. Assuming that
the thickness of the period is much less than light wavelength
$\lambda$ and using the effective medium approach, we get the
principle components of the effective medium permittivity tensor
\cite{Ferrari}
\begin{eqnarray}
\varepsilon_{xx} &=& \varepsilon_{yy} = p \varepsilon_{m}(\omega) +
(1-p)\varepsilon_{ad}, \label{epsxx} \\
\varepsilon_{zz} &=& \left( \frac{p}{\varepsilon_{m}(\omega)} +
\frac{1-p}{\varepsilon_{ad}} \right)^{-1}, \label{epszz}
\end{eqnarray}
where $p=d_m/(d_m+d_d)$ is the AZO filling factor (part of the
multilayer volume occupied by AZO), and $d_m$ and $d_d$ are the
thicknesses of the AZO and dielectric slabs, respectively. In Fig.
\ref{fig2} we show the real and imaginary parts of
$\varepsilon_{xx}$ and $\varepsilon_{zz}$  for $p=1/3$ and
$w_{eq}=0$ (i.e., $\varepsilon_{ad}=n_d^2$). The imaginary part of
$\varepsilon_{ad}$ does not significantly affect the effective
permittivities. In the wavelength range $\lambda \approx 2.2 - 3.1$
$\mu$m the real parts of the permittivity components have opposite
signs: ${\rm Re}\varepsilon_{xx}>0$ and ${\rm
Re}\varepsilon_{zz}<0$. This means that in this spectral range the
multilayer can be considered as an HMM of type I (dielectric type)
\cite{Ferrari}. The similar effective permittivity was reported for
the passive AZO/ZnO multilayer \cite{Naik12}. In agreement with the
Kramers-Kronig relations, absorption grows within this wavelength
range of the resonance as evidenced by the imaginary part of the
permittivity in Fig. \ref{fig2}(b). Our goal is to compensate for
absorption by embedding a gain medium into the multilayer.

We consider a subwavelength multilayer having $N=50$ periods. The
thicknesses of the AZO and dielectric slabs are $d_m=0.1$ $\mu$m and
$d_d=0.2$ $\mu$m, respectively, i.e. the filling factor equals
$p=1/3$, as in Fig. \ref{fig2}. We study temporal dynamics of a
monochromatic continuous wave (cw) at $\lambda=2.5$ $\mu$m.
Reflection $R$ and transmission $T$ coefficients depend on pumping
parameter $w_{eq}$ for small incident amplitude $\Omega_0=10^{-4}
\gamma_2 \ll \Omega_{sat}$ as demonstrated in Figs. \ref{fig3}(a)
and \ref{fig3}(b), respectively. In the gainless case, the
stationary reflection level is about $37 \%$, while the transmission
is less than $1 \%$. The rest energy of the incident wave is
absorbed by the structure. Turning on gain in the dielectric slabs
results in the gradual increase of both reflection and transmission
up to $1.38$ and $0.30$, respectively, at the maximum pumping level.
The oscillations of reflected and transmitted signals seen at
$w_{eq}=-1$ are due to the abrupt switching on of the incident wave
and can be eliminated with a smoothly rising input. The fact that we
observe a steady-state amplified output signal for every $w_{eq}$ in
Figs. \ref{fig3}(a) and \ref{fig3}(b) stems from the absence of
saturation. Then, according to Eq. (\ref{Nstat}), the population
difference remains unchanged. This is proved by direct computation
of $w(t)$ shown in the inset of Fig. \ref{fig3}(a): the population
difference keeps its initial value $w_{eq}=-1$ for all times.

When saturation effects are not negligible, the output signal is not
stationary despite the same strong amplification of the input signal
$\Omega_0=10^{-2} \gamma_2$ as before [see Figs. \ref{fig3}(c) and
\ref{fig3}(d)]. Reflection $R$ and transmission $T$ slowly degrade
to the level of the passive structure, because the input intensity
is able to change the population difference toward the saturated
value $w=0$ as clearly shown in the inset of Fig. \ref{fig3}(c).

So, calculations taking into account saturation demonstrate the
possibility of loss compensation and signal amplification in lossy
multilayers including HMMs. Albeit we are limited here to normal
incidence, we anticipate the similar loss compensation valid for
oblique incidence as well.

\section{\label{trans} Resonant enhanced transmission with suppressed reflection in AZO-dielectric multilayers}

In spite of simultaneous amplification of reflection and
transmission shown in Figs. \ref{fig3}(a) and \ref{fig3}(b), the
multilayer structure remains predominantly reflective in both
passive and active regimes, whereas for the practical applications
(e.g., lensing) the structure should be rather transmissive. In this
section, we demonstrate the suppressed reflection regime with
simultaneous enhancement of transmission. This effect takes place
within a narrow frequency band; hence it is resonant by its nature.

\begin{figure*}[t!]
\includegraphics[scale=0.95, clip=]{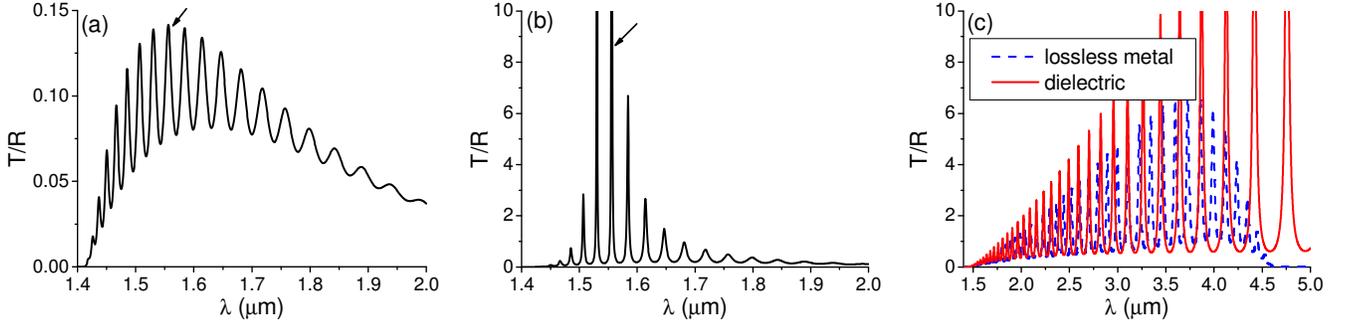}
\caption{\label{fig4} Ratio of the transmission to reflection
spectra of the AZO-dielectric structure for the pumping parameters
(a) $w_{eq}=0$ and (b) $w_{eq}=-0.2$. (c) The ratio $T/R$ in the
case of lossless ($\gamma = 0$) oxide-dielectric and entirely
dielectric ($\varepsilon_m = 3.45$) systems (the corresponding
curves are dashed blue and solid red) for the pumping parameter $w_{eq}=-0.2$.
Thicknesses $d_m=0.1$ $\mu$m and $d_d=0.12$ $\mu$m and refractive
index $n_d=3.4$, while the other parameters of calculations are the
same as in the caption of Fig.~\ref{fig2}. All calculations are
carried out using the transfer-matrix method.}
\end{figure*}

\begin{figure*}[t!]
\includegraphics[scale=0.9, clip=]{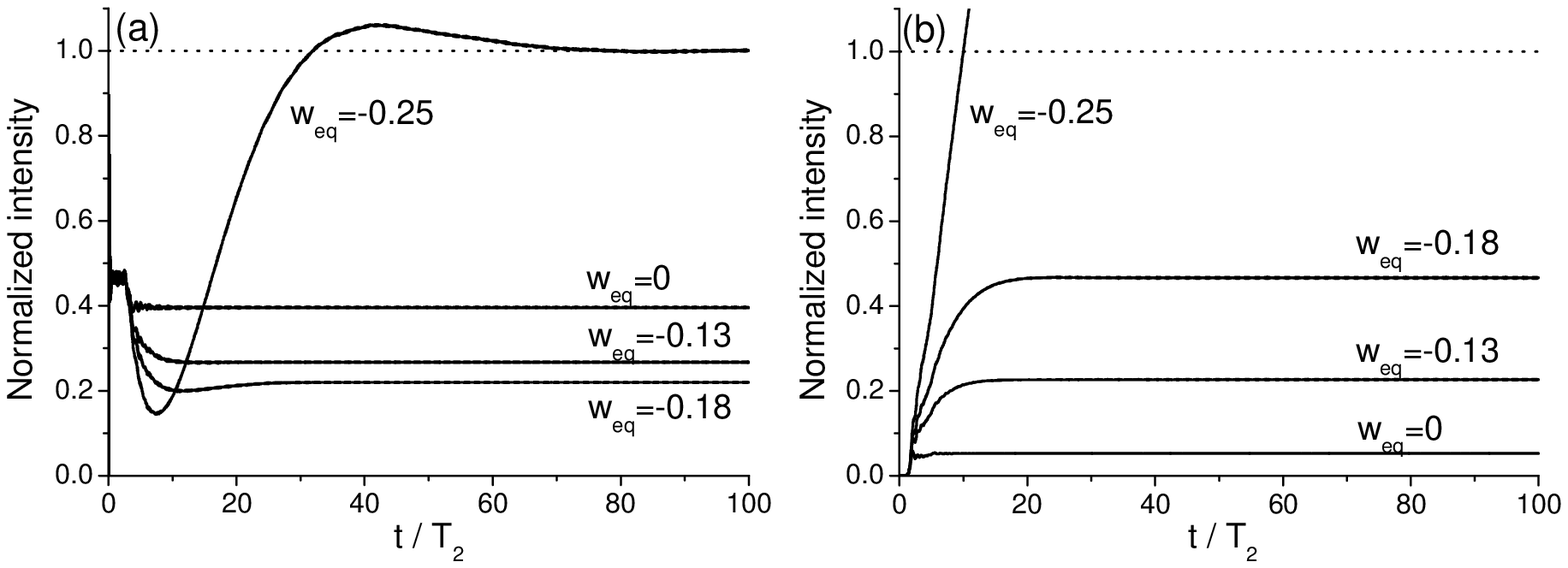}
\caption{\label{fig5} Temporal dynamics of (a) reflected and (b)
transmitted intensity for different pumping parameters. Wavelength
$\lambda=1.555$ $\mu$m (C-telecom band) corresponds to the peak marked with the arrow in
Fig.~\ref{fig4}(b). Parameters of calculations can be found in the
caption of Fig.~\ref{fig4}.}
\end{figure*}

We consider this regime in the structure with the same parameters as
in the previous section, except those of the dielectric slabs (we
take $d_d=0.12$ $\mu$m and $n_d=3.4$). Index $n_d$ is changed in
order to take into account the dispersion of the semiconductor
material \cite{Palik}, when the wavelength is within the
conventional telecommunications band in the vicinity of $1.55$
$\mu$m.

First, using the transfer-matrix method, we calculate reflection $R$
and transmission $T$ spectra of the structure with the slab
permittivities given by Eqs. (\ref{epsm}) and (\ref{epsTLMsimple}).
The ratio $T/R$ in Fig.~\ref{fig4}(a) clearly shows that reflection
dominates in passive structures. In the active system, transmission
takes over at the resonant wavelengths as demonstrated in
Fig.~\ref{fig4}(b) for $w_{eq} = -0.2$. Comparative analysis of
Figs.~\ref{fig4}(a) and (b) reveals the same number of peaks. This
allows us to conclude that the strong resonances are the amplified
seed peaks available in the gainless case. The seed peaks apparently
stem from the properties of the multilayer. If one substitutes the
multilayer with a homogeneous anisotropic material, the permittivity
of which is given by Eq.~(\ref{epszz}), the resonances move to other
wavelengths due to the inaccuracies inherent in the effective medium
approximation.

In the case of lossless oxide the number of amplified peaks
increases, but their intensity noticeably drops [compare dashed blue
lines in Fig.~\ref{fig4}(c) with the spectrum in Fig.~\ref{fig4}(b),
where the peak of the $T/R$ ratio is up to 70]. Therefore, the
conducting properties are crucial for the enhanced transmission,
since the loss results in redistributing energy to the few narrower
resonances. When the system is fully dielectric, i.e.,
$\varepsilon_m$ is not negative, the resonances providing the
amplified transmission at the expense of the reflection exist up to
long wavelengths [Fig.~\ref{fig4}(c)]. In the frequency domain, it
is clearly observed that the peaks are separated by the same
interval and correspond to the Fabry-P\'erot resonances between the
boundaries of the multilayer.

\begin{figure*}[t!]
\includegraphics[scale=0.9, clip=]{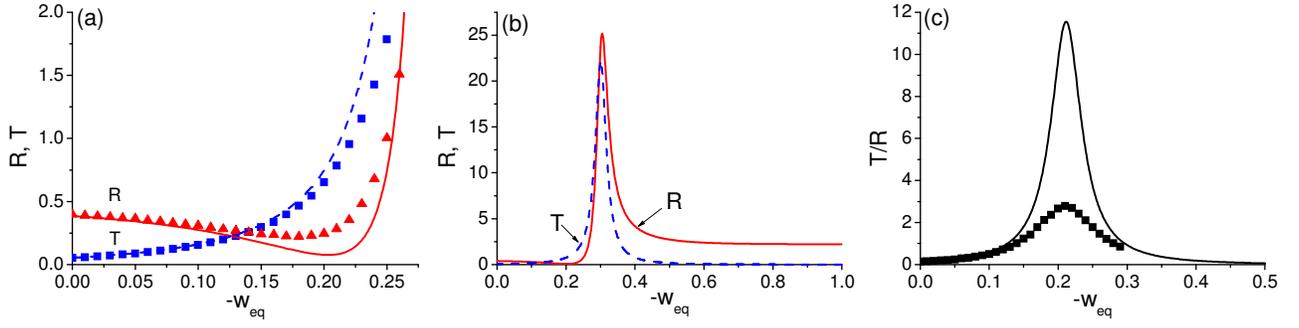}
\caption{\label{fig6} Dependence of the stationary level of
reflection (R) and transmission (T) on the pumping parameter
$w_{eq}$ at $\lambda=1.555$ $\mu$m. (a) Results of numerical
simulations (symbols) and transfer-matrix calculations (lines). (b)
Transfer-matrix results in the full range of pumping parameter. (c)
Transmission to reflection ratio from the numerical
simulations (symbols) and transfer-matrix calculations (line).}
\end{figure*}

\begin{figure*}[t!]
\includegraphics[scale=0.9, clip=]{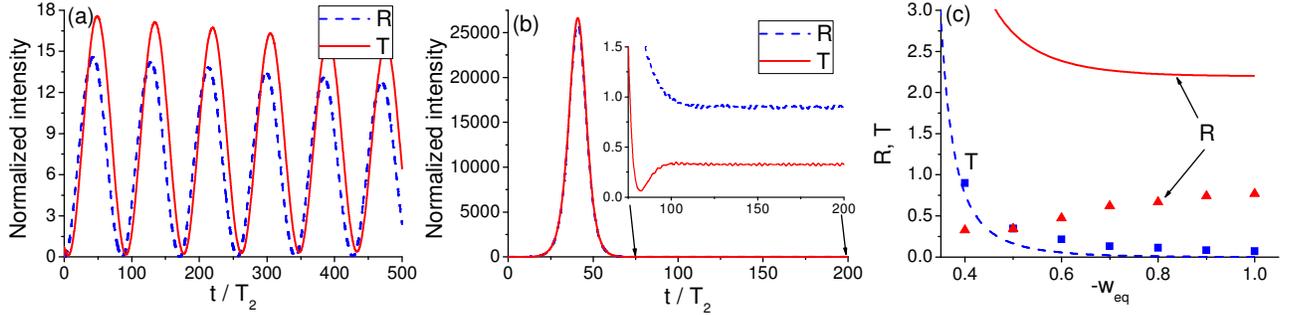}
\caption{\label{fig7} (a) Temporal dynamics of the reflected (R) and
transmitted (T) intensity for the pumping
parameter $w_{eq}=-0.3$. (b) The same as in (a), but for $w_{eq}=-0.4$.  (c)
Dependence of the stationary level of reflection (R) and
transmission (T) on the pumping parameter for the structure in the
lasing-like mode derived from the numerical
simulations (symbols) and transfer-matrix calculations (lines). Parameters of calculations are the same as in the
caption of Fig.~\ref{fig4}.}
\end{figure*}

We take wavelength $\lambda=1.555$ $\mu$m of the incident wave
corresponding to the transmission peak [see Fig.~\ref{fig4}(b),
where this resonance is marked with the arrow]. The transfer-matrix
method estimates decrease in reflection from $0.38$ to $0.07$ and
increase in transmission from $0.05$ to $0.75$ as pumping parameter
$-w_{eq}$ grows from $0$ to $0.2$. Since the spectra in
Fig.~\ref{fig4} provide only a qualitative stationary picture, we
simulate propagation taking into account nonuniformity of the light
field across the layers and saturation of the gain medium response.
To suppress the effect of saturation, the input wave amplitude is
chosen to be $\Omega_0=10^{-4} \gamma_2$. After establishment of the
stationary regime, the passive structure reflects about $40 \%$ of
incident light and transmits almost nothing (Fig.~\ref{fig5}). This
is in full accordance with the transfer-matrix calculations. For the
greater $-w_{eq}$, the establishment of the stationary regime takes
more time, reflection gradually decreases, and, on the contrary,
transmission increases. They are approximately equal at
$w_{eq}=-0.13$ as demonstrated by the corresponding curves in
Fig.~\ref{fig5}. The reflection coefficient reaches its minimal
stationary value at $w_{eq}=-0.18$ with the transmission larger than
$45 \%$ [Fig.~\ref{fig5}(b)]. Higher pumping leads to the rapid
growth of both reflection and transmission coefficients. At $w_{eq}
\approx -0.3$, the system switches to another operation mode, in
which the stored pumped energy is promptly released in the form of a
powerful pulse. This regime, that can be called the lasing-like
mode, will be addressed afterwards.

The stationary levels of reflection $R$ and transmission $T$
depending on pumping parameter $w_{eq}$ are depicted in
Fig.~\ref{fig6}(a). There is a good agreement between the results
obtained from direct numerical simulations (symbols) and
transfer-matrix calculations (lines). Intersection of the reflection
and transmission curves at $w_{eq} \approx -0.13$ separates two
regimes (reflective and transmissive) of the system. The optimal
conditions for reflecting and transmitting states are realized for
small pumping parameters and for $w_{eq} \approx -0.2$,
respectively. There is a mixture of the states near the crossing
point of the reflection and transmission curves, and the transition
of the structure from the reflective regime to the transmissive one
occurs. It should be noted that the stationary transfer-matrix
calculations not only specify the perspective frequency band
containing the resonance, but also correctly predict the
intersection point and optimal pumping for minimal reflection. On
the other hand, this method cannot describe peculiarities of the
transition between two states and generally overestimates the depth
of the reflection minimum as clearly shown in Fig.~\ref{fig6}(a). We
should note that almost identical dependencies as shown in this
figure can be obtained for the total reflection and transmission
energies of incident pulses with durations long enough to be
considered as quasimonochromatic. Figure~\ref{fig6}(b) shows the
transfer-matrix results for the stationary level of reflection and
transmission in the full range of pumping parameters $w_{eq}$. The
peaks in the vicinity of $w_{eq} \approx -0.3$ can be treated as a
qualitative indication of the lasing-like mode, though the
quantitative characteristics of this regime can be revealed only by
solving the full system of equations described in Sec. \ref{eqpars}.
The optimal pumping parameter for the transmissive state of the
system extracted from the transmission-to-reflection ratio [solid
line in Fig.~\ref{fig6}(c)] well correlates with the numerical
simulations (symbols), but the value of this ratio is strongly
overestimated by the transfer-matrix method. This can be explained
by the detuning of the resonant frequency in the FDTD approach as
compared with the transfer-matrix method.

For relatively low pumping parameters $-w_{eq}<0.28$ as in
Fig.~\ref{fig6}(a) there is no saturation, and population difference
$w$ in the gain medium does not vary with time. Even at the edge
value $w_{eq}=-0.3$ the population difference is time independent,
while the reflected and transmitted signals in Fig.~\ref{fig7}(a)
demonstrate slowly decaying oscillations, i.e., there is a very long
transient period in this case. Therefore, we cannot assign any
steady-state value to reflection and transmission for the
intermediate parameter range $-0.35 \lesssim w_{eq} \lesssim -0.28$.
Nevertheless, when the pumping crosses over a certain threshold
level, saturation comes into play. The temporal dependence of the
population difference starts as an accrescent fluctuation analogous
to that shown in the inset of Fig.~\ref{fig3}(c), while the
amplified signal takes the form of a pulse with the peak intensity
exceeding the input cw intensity in hundreds or even thousands
times. The example for $w_{eq}=-0.4$ shown in Fig.~\ref{fig7}(b)
demonstrates almost identical pulses in both directions. The pulse
duration shortens for higher pumping. This phenomenon of powerful
pulse generation differs from the usual lasing action, because the
response develops not from fluctuations of the light field inside
the structure, but rather from the strong external radiation. This
is the reason why we call this regime the lasing-like mode. We
should also emphasize that the transition to this mode has the
resonant character (compare with Fig. \ref{fig3} where lasing-like
regime is absent in the full range of pumping) and has approximately
the same threshold for resonances in different frequency ranges.

Pulsed-form amplification is followed by formation of the stationary
response after the pumped energy has been emitted. At $w_{eq}=-0.4$,
transmission is higher than reflection, they are almost equal at
$w_{eq}=-0.5$, and then reflection (transmission) increases
(decreases) keeping absorption level approximately at the same level
of $0.2$ [see symbols in Fig.~\ref{fig7}(c)]. The structure becomes
almost entirely reflective in the large-time limit at $w_{eq}=-1$.
We should emphasize that the system does not return to the gainless
state, since reflection is much higher than that found in the
passive system. Note that the transfer-matrix calculations give
strongly overestimated values of the stationary level of reflection
[see lines in Fig.~\ref{fig7}(c)], because the inhomogeneous field
distribution and saturation within the gain layers cannot be ignored
in this case.

\section{Resonant reflectionless transmission for obliquely incident and evanescent waves} \label{oblique}

\begin{figure*}[t!]
\includegraphics[scale=1.6, clip=]{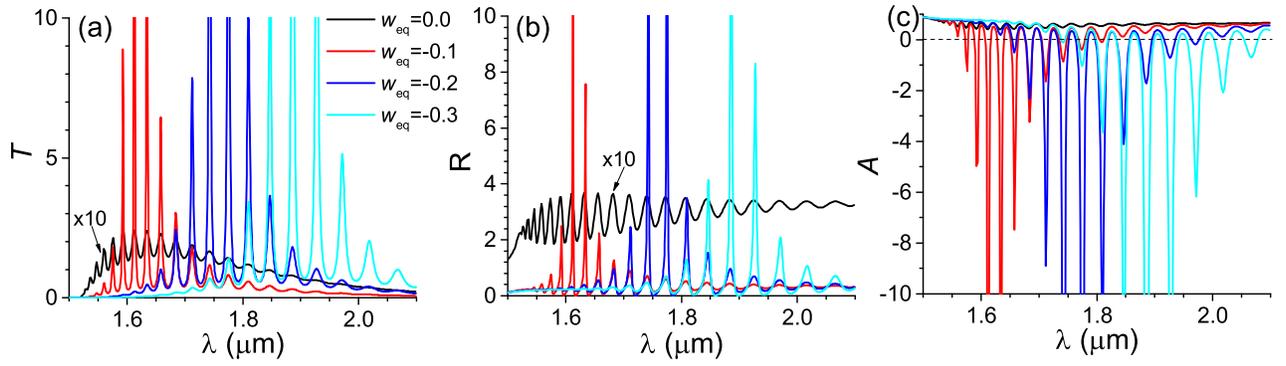}
\caption{\label{fig8} (a) Transmission $T$, (b) reflection $R$, and
(c) absorption $A = 1-T-R$ for the TE-polarized evanescent wave
($k_x = 2 k_0$) propagating through the passive and active
multilayers ($d_m = 100$ nm and $d_d = 200$ nm). Parameters of
calculations are the same as in Fig.~\ref{fig2}.}
\end{figure*}

\begin{figure*}[t!]
\includegraphics[scale=0.9, clip=]{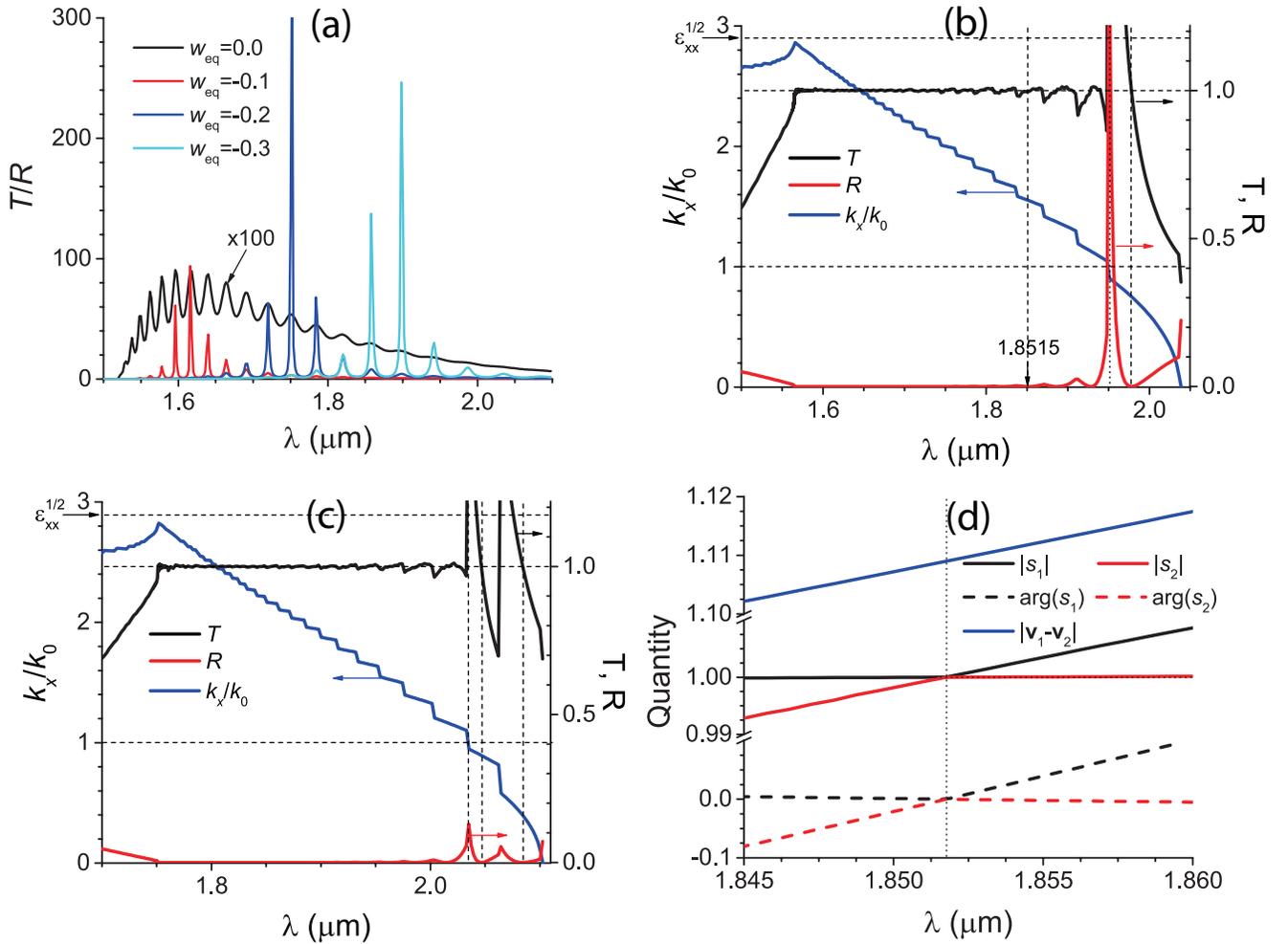}
\caption{\label{fig9} (a) Transmission [Fig. \ref{fig8}(a)] to
reflection [Fig. \ref{fig8}(b)] ratio. Dispersion equation
$k_x(\lambda)$ of the peak value of the ratio $T/R$ (left scale) and
transmission $T$ and reflection $R$ spectra at this peak (right
scale) for TE-polarized waves in the active multilayers ($d_m = 100$
nm and $d_d = 200$ nm): (b) $w_{eq}=-0.2$ and (c) $w_{eq}=-0.3$. (d)
Eigenvalues $s_1$ and $s_2$ and difference of eigenvectors $|{\bf
v}_1-{\bf v}_2|$ of scattering matrix near the point of $T=1$ and
$R=0$ [$\lambda \approx 1.8515$ in (b)]. Other parameters of
calculations are the same as in Fig.~\ref{fig2}.}
\end{figure*}

\begin{figure}[t!]
\includegraphics[scale=0.7, clip=]{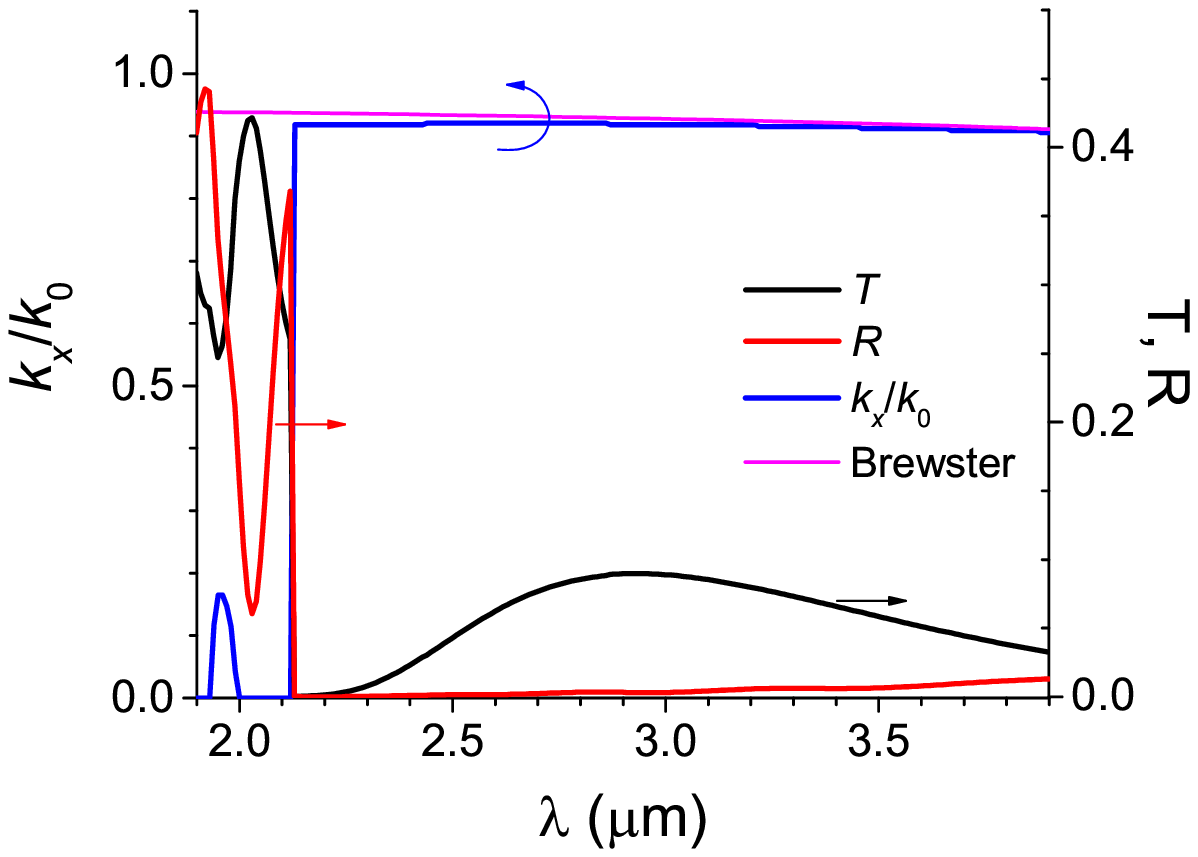}
\caption{\label{fig10} Dispersion equation $k_x(\lambda)$ of the
peak value of the ratio $T/R$ (left scale) and transmission $T$ and
reflection $R$ spectra at this peak (right scale) for TM-polarized
waves in the active metal-dielectric multilayer with $d_m = 20$ nm,
$d_d = 40$ nm, and $w_{eq}=-0.2$. Other parameters of calculations
are the same as in the caption of Fig.~\ref{fig2}.}
\end{figure}

An obliquely incident wave is characterized by transverse $k_x$ and
longitudinal $k_z$ wavenumbers directed along and across the layers,
respectively. Transverse component is conserved at the multilayer
interfaces being defined by  incident angle $\alpha$ of the input
wave as follows: $k_x = k_0 \sin\alpha$ (the ambient medium is free
space). If an evanescent wave is excited at the input, then
transverse wavenumber $k_x > k_0$, while longitudinal wavenumber
$k_z$ is defined by the corresponding dispersion equation.

Transmission of the obliquely incident and evanescent waves is
sensitive to their polarization, which can be easily noticed within
the effective medium approximation [see Eqs.~(\ref{epsxx}) and
(\ref{epszz})]. TE-polarized plane waves in such a homogenized
medium are the ordinary waves of the effective uniaxial crystal with
the $z$-oriented optical axis. Their dispersion is governed by the
equation for isotropic materials: $k_x^2 + k_z^2 = \varepsilon_{xx}
k_0^2$. Since ${\rm Re}(\varepsilon_{xx}) >0$ (see Fig.~\ref{fig2}),
propagation of waves with longitudinal wavenumbers $k_x$ up to $k_0
\sqrt{{\rm Re}(\varepsilon_{xx})}$ is allowed. Dispersion equation
of TM-polarized (extraordinary) plane waves reads
$k_x^2/\varepsilon_{zz} + k_z^2/\varepsilon_{xx} = k_0^2$. If ${\rm
Re} (\varepsilon_{zz})>0$, the tangential wavenumber cannot be
greater than $k_0 \sqrt{{\rm Re}(\varepsilon_{zz})}$. Otherwise,
when ${\rm Re} (\varepsilon_{zz})<0$, the structure can, in
principle, sustain an evanescent wave with any $k_x$.

In this section, the calculations are carried out using the
transfer-matrix method. According to Fig. \ref{fig8}, the peaks of
active and passive structures agree with each other for the
evanescent waves. The system exhibits resonant amplification for
each pumping $w_{eq}$, transmission maximums being generally
stronger than reflection ones. Out of the resonances the system is
strongly absorbing [see Fig. \ref{fig8}(c)]. Points of low
reflection result in the maximums of $T/R$. Figure~\ref{fig9}
clearly demonstrates that the evanescent waves can be amplified
exclusively in transmission. Indeed, similar to the case of normal
incidence (Fig. \ref{fig4}), ratio $T/R$ is small for the passive
structure and strongly enhanced for active multilayers [see
Fig.~\ref{fig9}(a)]. The greater is the pumping parameter, the
larger is the redshift of the major peak. Efficiency of
amplification apparently depends not only on $w_{eq}$, but also on
the spectral characteristics of the passive structure. In order to
study the features of the waves at the peaks of $T/R$, we find
transverse wavenumbers $k_x$ of the waves corresponding to one of
such peaks as a function of wavelength $\lambda$. In other words,
there is a correspondence between the position of the peak $\lambda$
and wavenumber $k_x$ which provides the maximum of the $T/R$ ratio.
The procedure of retrieving dispersion $k_x(\lambda)$ is as follows.
Transmission and reflection depend on $\lambda$ and $k_x$. For each
wavelength $\lambda$, we introduce wavenumbers $k_x^{(j)}$
($j=1,\ldots,M$) in the chosen region, calculate
$T(\lambda,k_x^{(j)})/R(\lambda,k_x^{(j)})$, and determine the
wavenumber corresponding to the maximal ratio. Thus we have the pair
($\lambda$, $k_x$). The whole set of such pairs results in the
dispersion curve $k_x(\lambda)$. Then we calculate transmission $T$
and reflection $R$ coefficients for the values lying on the
dispersion curve. A slight jitter of the values in
Fig.~\ref{fig9}(b) appears due to inaccuracy of finding the peak
values of the narrow resonances $T/R$. The strongest jitter occurs
near $k_z = 0$ [wavelength is about $1.6$ $\mu$m in Fig.
\ref{fig9}(b)]. To find the correct positions of the narrow peaks in
this region and, therefore, suppress the deviations, we increase the
accuracy of our calculation (number $M$ of discrete wavenumbers) in
100 times there.

When $k_x(\lambda) < k_0$, we are in the realm of propagating waves.
In this regime, reflection completely vanishes and transmission is
unity for a single obliquely incident wave with $k_x \approx 0.75
k_0$ (the angle of incidence is $\alpha_0 = 48.5^\circ$), which
corresponds to the wavelength of the peak around $1.98$ $\mu$m [see
the vertical dotted line in Fig. \ref{fig9}(b)]. Both $T$ and $R$
rapidly grow for angles $\alpha > \alpha_0$. Enormous values of
transmission and reflection resemble those observed in Fig.
\ref{fig6}(b). The transfer-matrix method is likely to be
inapplicable for quantitative conclusions, but it should work quite
well in the regime of high transmission with suppressed reflection.
If $\alpha < \alpha_0$, transmission drops and becomes equal to
reflection near $k_x = 0$ [see the far right part of Fig.
\ref{fig9}(b)]. It is interesting that the incident evanescent waves
($k_x>k_0$) can result in higher $T/R$ than the propagating waves.
Moreover, the regime of high transmission with well suppressed
reflection is observed in the wide spectral range and ends at $k_x =
k_0 \sqrt{{\rm Re}\varepsilon_{xx}}$. However, full transmission
with no reflection exists only at the specific wavelengths. When
pumping parameter $w_{eq}$ decreases (increases), the picture in
Fig.~\ref{fig9}(b) red (blue) shifts, because the higher gain is
able to compensate the higher multilayer losses shown in
Fig.~\ref{fig2}(b). Behavior of the system at oblique incidence
($k_x < k_0$) may also change, e.g., revealing several wavelengths
of full transmission as demonstrated in Fig.~\ref{fig9}(c). When the
thicknesses of the AZO and dielectric slabs are enlarged, we
approach the photonic crystal mode, and the moderate oscillations
observed in transmission of the evanescent waves in
Figs.~\ref{fig9}(b) and \ref{fig9}(c) become more pronounced.

To achieve reflectionless transmission, we do not impose any
specific condition on the system in question, although the full loss
compensation can be forbidden in the quasistatic consideration of
plasmonic structures \cite{Stockman11}. On the other hand,
$\mathcal{PT}$-symmetric loss-gain systems demonstrate transmission
without reflection \cite{Shramkova}.

Let us look closer at the behavior of the system near the points of
full transmission. For instance, both conditions $T=1$ and $R=0$ are
met in the vicinity of wavelength $\lambda = 1.8515$ $\mu$m [see
Fig.~\ref{fig9}(b)]. Since reflectionless transmission is known for
$\mathcal{PT}$-symmetric structures, it is instructive to compare
them with the system under consideration. We calculate the
scattering matrix $S$ and similar to $\mathcal{PT}$-symmetric
structures \cite{Ge12}, we obtain $s_1=s_2=1$  at the point
characterized by $T=1$ and $R=0$ [see the absolute values and
arguments for complex eigenvalues in Fig.~\ref{fig9}(d)]. However,
our system only resembles the $\mathcal{PT}$-symmetric one at the
exceptional point, because, first, it does not have the necessary
distribution of permittivity $\varepsilon(z) = \varepsilon^*(-z)$
and, second, the eigenvectors of the scattering matrix are not the
same at this point [see their difference in Fig.~\ref{fig9}(d)].
Since the system is not $\mathcal{PT}$ symmetric, the condition of
symmetry-broken state $|s_2|= 1/|s_1|$ is not valid, too. We believe
that the emergence of the points of reflectionless transmission is
caused by resonant properties of the multilayers.

As demonstrated in Fig. \ref{fig2}, there is a spectral range of
hyperbolic dispersion from $2.2$ $\mu$m to $3.1$ $\mu$m available
for the TM-polarized waves. However, the resonant reflectionless
transmission is not observed for these waves. The suppressed
reflection in this case is caused by the Brewster effect as shown in
Fig. \ref{fig10} [compare the curve for the Brewster effect in the
effective medium approximation with curve $k_x(\lambda)$].
Evanescent waves possess a much lower $T/R$ ratio compared to the
propagating waves and, hence, are absent in the dispersion curve of
Fig. \ref{fig10}.

\section{\label{concl}Conclusion}

The system of alternating transparent conducting oxide (AZO) and
amplifying dielectric (semiconductor doped with quantum dots) slabs
has been studied in two modes: loss compensation and resonant
low-reflection transmission. Using the FDTD method, we have shown
that the stationary regime is rapidly established giving us the
opportunity to exploit the transfer-matrix approach for the
realistic description of the system. In the loss compensation mode,
we have analyzed temporal evolution of the reflected and transmitted
signals with and without saturation demonstrating that saturation
suppresses signal amplification in the long-time limit. Harnessing
the resonant properties of multilayers, we have found the conditions
for transmission enhancement with simultaneous suppression of
reflection. The resonant wavelengths can be estimated within the
transfer-matrix approach for stationary fields. Temporal dynamics
qualitatively confirms existence of the maximum in the ratio of
transmission to reflection, but the specific value of $T/R$ differs.
Since the resonance is narrow, this difference is likely caused by a
small deviation of the resonant frequency in FDTD and
transfer-matrix calculations. We have revealed the lasing-like mode
at large enough pumping, the appearance of which is again associated
with the multilayer resonances. In this mode, both transmitted and
reflected fields are released as powerful pulses with subsequent
formation of the stationary levels of reflection and transmission,
which cannot be predicted correctly by the transfer-matrix approach.

Interaction of the obliquely incident and evanescent waves with the
multilayer has been studied using the transfer-matrix method. We
have found a number of discrete points, where the reflection equals
zero, while the transmission is unity. Although the system at these
points may resemble a $\mathcal{PT}$-symmetric one (the eigenvalues
of the scattering matrix are equal), it is not the case, because the
permittivity condition $\varepsilon(z) = \varepsilon^*(-z)$ is not
satisfied and the eigenvectors of the scattering matrix are not the
same. We guess that emergence of reflectionless transmission stems
rather from the resonant properties of the multilayer, in which the
Bloch waves can propagate without amplification or attenuation. It
should be noted that the $\mathcal{PT}$-symmetric-like behavior has
been also observed in dissimilar directional couplers
\cite{Walasik17}. In our paper, the reflectionless transmission has
been found only for TE-polarized obliquely incident and evanescent
waves, but this may not be the rule. To conclude, we have revealed
the mechanism of total transmission of evanescent waves by means of
their conversion into propagating modes of the multilayer.
Reflectionless transmission for evanescent waves opens new
opportunities for the light-matter interaction in plasmonics and
metamaterials.

\acknowledgements{This work was supported by the Belarusian
Republican Foundation for Fundamental Research (Projects No.
F16K-016 and No. F16R-049) and the State Fund for Fundamental
Research of Ukraine (Project No. F73). Numerical simulations of
light interaction with resonant media were supported by the Russian
Science Foundation (Project No. 17-72-10098). Partial financial
support from the Villum Fonden via the DarkSILD project and from the
HyMeCav project is acknowledged. }

\end{document}